\documentclass[11pt]{article}
\usepackage{graphicx}
\begin{document}

\begin{center}
{\bf\large Coherent states for polynomial su(1,1) algebra and a
conditionally solvable system}
~\\[5mm]

Muhammad Sadiq\\[3mm]

National Center for Physics\\

Quaid-i-Azam University\\

Islamabad 45320, Pakistan\\[5mm]

Akira Inomata\\[3mm]

Department of Physics\\

State University of New York at Albany\\

Albany, New York 12222, USA\\[5mm]

Georg Junker\\[3mm]

European Organization for Astronomical Research in the Southern
Hemisphere\\

Karl-Schwarzschild-Strasse 2, D-85748 Garching, Germany \\[5mm]

\end{center}

{\bf Abstract} In a previous paper [{\it J. Phys. A: Math. Theor.} {\bf
40} (2007) 11105], we constructed a class of coherent states for a
polynomially deformed $su(2)$ algebra. In this paper, we first prepare
the discrete representations of the nonlinearly deformed $su(1,1)$
algebra. Then we extend the previous procedure to construct a discrete
class of coherent states for a polynomial $su(1,1)$ algebra which
contains the Barut-Girardello set and the Perelomov set of the $SU(1,1)$
coherent states as special cases. We also construct coherent states for
the cubic algebra related to the conditionally solvable radial oscillator
problem.

\section{Introduction}

In a previous paper \cite{SI}, we have constructed a set of coherent
states for a polynomially deformed $su(2)$ algebra. The goal of the
present paper is to construct a discrete class of coherent states for a
polynomial $su(1,1)$ algebra by extending the procedure employed for the
polynomial $su(2)$ case. For the usual $SU(1,1)$ group, there are two
well-known sets of coherent states: the Barut-Girardello coherent states
\cite{BG} which are characterized by the complex eigenvalues $\xi $ of
the noncompact generator $\hat{K}_{-}$ of the $su(1,1)$ algebra
\begin{equation}
\hat{K}_{-}|\xi \rangle = \xi |\xi \rangle , \label{BG}
\end{equation}
and the Perelomov coherent states \cite{Pere} which are characterized by
points $\eta $ of the coset space $SU(1,1)/U(1)$
\begin{equation}
|\eta \rangle = N^{-1}\,e^{\eta \hat{K}_{+}}|0\rangle, ~~~~~~\hat{K}_{-
}|0\rangle =0. \label{P}
\end{equation}
These two sets are not equivalent. Since we have no knowledge of the
group structure corresponding to the polynomial $su(1,1)$ algebra, we
are unable to follow Perelomov's group theoretical approach. Thus we
construct coherent states in such a way that they are reducible either
to the Barut-Girardello $SU(1,1)$ states or the Perelomov $SU(1,1)$
states in the linear limit. In the literature \cite{Cha,CJT,SBJP},
several authors have proposed various sets of coherent states for the
polynomial $su(1,1)$ algebra in different contexts. What we wish to
study here is a unified treatment of coherent states of the
Barut-Girardello type and the Perelomov type for the polynomial
$su(1,1)$, which differs from all of those reported earlier.

The polynomial $su(2)$ algebra we considered earlier \cite{SI} is a
special case of the nonlinearly deformed $su(2)$ algebra of Bonatos,
Danskaloyannis and Kolokotronis (BDK) \cite{BDK}. BDK's deformed
algebra, denoted by $su_{\Phi }(2)$, is of the form,
\begin{equation}
[\hat{J}_{0}, \hat{J}_{\pm}]=\pm \hat{J}_{\pm}, ~~~~~
[\hat{J}_{+}, \hat{J}_{-}] = \Phi
\left(\hat{J}_{0}(\hat{J}_{0}+1)\right)
- \Phi \left(\hat{J}_{0}(\hat{J}_{0}-1)\right), \label{sup2}
\end{equation}
where the structure function $\Phi (x)$ is an increasing function of
$x$ defined for $x \geq -1/4$. The Casimir operator for
$su_{\Phi }(2)$ is
\begin{equation}
\hat{\bf J}^{2}=\hat{J}_{-}\hat{J}_{+} + \Phi \left(\hat{J}_{0}(\hat{J}_{0}
+1)\right) =\hat{J}_{+}\hat{J}_{-} + \Phi \left(\hat{J}_{0}(\hat{J}_{0}
-1)\right). \label{C}
\end{equation}
On the basis $\{|j, m\rangle \}$ that diagonalizes $\hat{\bf J}^{2}$ and
$\hat{J}_{0}$ simultaneously such that \cite{BDK}
\begin{equation}
\hat{\bf J}^{2}|j,m\rangle =\Phi \left(j(j+1)\right)|j, m\rangle ,
~~~~~~~~~\hat{J}_{0}|j, m\rangle = m |j, m\rangle  , \label{su2base}
\end{equation}
the operators $\hat{J}_{+}$ and $\hat{J}_{-}$ satisfy the relations
\begin{equation}
\hat{J}_{+}|j, m\rangle = \sqrt{\Phi \left(j(j+1)\right) - \Phi
\left(m(m+1)\right)}\,|j, m+1 \rangle
\end{equation}
\begin{equation}
\hat{J}_{-}|j, m\rangle = \sqrt{\Phi \left(j(j+1)\right) - \Phi
\left(m(m-1)\right)}\,|j, m-1 \rangle
\end{equation}
with $2j=0,1,2,\cdots,$ and $|m| \leq j.$

The coherent states we constructed for $su_{\Phi }(2)$ by letting
$m=-j+n ~(n= 0, 1, 2,..., 2j)$ were
\begin{equation}
|j, \xi \rangle = N_{\Phi }^{-1}(|\xi |)\,\sum_{n=0}^{2j}
\frac{\sqrt{[k_{n}]!} }{n!}\xi ^{n}|j, -j + n\rangle .  \label{CS1}
\end{equation}
Here
\begin{equation}
k_{n}=\Phi \left(j(j+1)\right) - \Phi \left((j-n)(j-n+1)\right),
\label{kn}
\end{equation}
and
\begin{equation}
[k_{n}]!=\prod_{j=1}^{n}k_{n}, ~~~~~~[k_{0}]!=1.
\end{equation}
The normalization factor was given by
\begin{equation}
N_{\Phi }^{2}(|\xi |)=\sum_{n=0}^{2j}\frac{[k_{n}]! \,|\xi
|^{2n}}{(n!)^{2}}. \label{norm}
\end{equation}

For our polynomial $su(2)$ case, we imposed the polynomial condition that
\begin{equation}
\Phi (x)=\sum_{r=1}^{p}\alpha _{r}\,x^{r} ~~~~~(\alpha _{r} \in {\bf R})
\label{phi}
\end{equation}
with $\alpha _{p} \neq 0$. We showed that the coherent states we
obtained include the usual $su(2)$ coherent states and the cubic $su(2)$
coherent states as special cases.

In the present paper, we first extend BDK's $su_{\Phi }(2)$ to a
nonlinearly deformed $su(1,1)$ algebra and prepare discrete
representations for the algebra which correspond to those belonging to
the positive discrete series of the irreducible unitary representations
of $SU(1,1)$. Then we construct formally a set of coherent
states for the deformed algebra $su_{\Phi}(1,1)$ by generalizing the
$SU(1,1)$ group element used for the Perelomov states. As before, we
also impose the polynomial condition (\ref{phi}) to specify the coherent
states for the polynomially deformed algebra $su_{2p-1}(1,1)$. Out of
the formal states so constructed, we select two sets of states which are
reducible to the Barut-Girardello set and the Perelomov set in the
linear limit. Finally we reformulate the conditionally solvable radial
oscillator problem in broken supersymmetric quantum mechanics, proposed
by Junker and Roy \cite{JR}, in an algebraic manner to show that the
eigenstates of one of the partner Hamiltonians, $\hat{H}_{+}$, in SUSY
quantum mechanics can be identified with a standard basis of the
$su(1,1)$ algebra whereas the set of eigenstates of the other partner
Hamiltonian $\hat{H}_{-}$ are identified with a representation space of
the cubic algebra $su_{3}(1,1)$. We also construct coherent states of
the Barut-Girardello type and of the Perelomov type for the
conditionally solvable problem.

\section{Polynomial $su(1,1)$ algebra and its representations}

In order to introduce a nonlinearly deformed $su(1,1)$ algebra in a
manner parallel to the nonlinearly deformed algebra $su_{\Phi }(2)$ of
Bonatos, Danskaloyannis and Kolokotronis \cite{BDK}, we exercise analytic
continuation \cite{BF,HB,IKG} on $su_{\Phi }(2)$. Replacing the
generators of $su_{\Phi }(2)$ in (\ref{sup2}) as
\begin{equation}
\hat{J}_{0} \rightarrow  \hat{K}_{0}, ~~~~~~~
\hat{J}_{\pm} \rightarrow  i\hat{K}_{\pm},  \label{repl}
\end{equation}
we extend $su_{\Phi }(2)$ formally into a deformed $su(1,1)$ algebra,
\begin{equation}
[\hat{K}_{0}, \hat{K}_{\pm}]=\pm \hat{K}_{\pm}, ~~~~~~~
[\hat{K}_{+}, \hat{K}_{-}] = \Phi
\left(\hat{K}_{0}(\hat{K}_{0}-1)\right)
- \Phi \left(\hat{K}_{0}(\hat{K}_{0}+1)\right), \label{sup11}
\end{equation}
which we denote by $su_{\Phi }(1,1)$ as an extension of BDK's $su_{\Phi
}(2)$. Here we assume that the generators of $su_{\Phi }(1,1)$ in
(\ref{sup11}) possess the hermitian properties,
\begin{equation}
\hat{K}_{0}^{\dagger}=\hat{K}_{0}, ~~~~~~~
\hat{K}_{\pm}^{\dagger}=\hat{K}_{\mp}. \label{hermi}
\end{equation}
We also assume that the structure function $\Phi (x)$ is a
differentiable function increasing with a real variable $x \geq -1/4$,
and is operator-valued and hermitian when $x$ is a hermitian operator.
Accordingly the operator obtainable from the Casimir operator (\ref{C})
of $su_{\Phi }(2)$ by the analytic continuation (\ref{repl}),
\begin{equation}
\hat{\bf K}^{2}=-\hat{K}_{-}\hat{K}_{+} + \Phi\left(\hat{K}_{0}(\hat{K}_{0}
+1)\right)= -\hat{K}_{+}\hat{K}_{-} + \Phi\left(\hat{K}_{0}(\hat{K}_{0}
-1)\right), \label{Q}
\end{equation}
is hermitian. From the first equation of (\ref{sup11}) immediately follows
\begin{equation}
\hat{K}_{0}^{r}\hat{K}_{\pm} = \hat{K}_{\pm}(\hat{K}_{0} \pm 1)^{r}
\label{Kmap}
\end{equation}
for $r = 0, 1, 2, ...$
Since the structure function $\Phi (x)$, assumed to be a real
differentiable function, can be expanded as a MacLaurin series, it is
obvious that
\begin{equation}
\Phi \left(\hat{K}_{0} (\hat{K}_{0} \mp 1)\right) \hat{K}_{\pm}
=\hat{K}_{\pm}\Phi \left(\hat{K}_{0} (\hat{K}_{0} \pm 1)\right).
\end{equation}
Therefore the operator $\hat{\bf K}^{2}$ of (\ref{Q}), being commutable with all
the three generators, is indeed the Casimir invariant of $su_{\Phi
}(1,1)$.

By imposing the polynomial condition (\ref{phi}) on the structure
function in (\ref{sup11}), we obtain a polynomial $su(1,1)$ algebra,
\begin{equation}
[\hat{K}_{0}, \hat{K}_{\pm}]=\pm \hat{K}_{\pm}, ~~~~~
[\hat{K}_{+}, \hat{K}_{-}] = - 2 \sum_{r=1}^{p}\alpha
_{r}\hat{K}_{0}^{r}\,\sum_{s=1}^{r}(\hat{K}_{0} + 1)^{r-s}(\hat{K}_{0} -
1)^{s-1}.     \label{supol}
\end{equation}
When $\Phi (x)$ is a polynomial in $x=\hat{K}_{0}(\hat{K}_{0} +1)$ of
degree $p$, the right hand side of the second equation of (\ref{supol})
becomes a polynomial in $\hat{K}_{0}$ of degree $2p-1$. Thus
(\ref{supol}) is the polynomial $su(1,1)$ algebra of odd degree $2p-1$
$(p=1,2,3,...)$, which we denote by $su_{2p-1}(1,1)$. As special cases,
$p=1$ and $p=2$ correspond to the usual $su(1,1)$ and the cubic algebra
$su_{cub}(1,1)$, respectively. The present scheme cannot accommodate
polynomial $su(1,1)$ algebras of even degree.

In analogy to the case of $su_{\Phi }(2)$ represented on the basis
$\{|j, m\rangle \}$ as in (\ref{su2base}), we consider a representation
space for $su_{\Phi }(1,1)$ which is spanned by simultaneous eigenstates
$\{|k, m\rangle \}$ of the Casimir operator $\hat{\bf K}^{2}$ and the compact
operator $\hat{K}_{0}$. On the basis $\{|k, m\rangle \}$, let $\hat{\bf
K}^{2}$ and $\hat{K}_{0}$ be diagonalized as
\begin{equation}
\hat{\bf K}^{2}|k, m \rangle = \Phi (k(k-1))|k, m \rangle ,
~~~~~~~\hat{K}_{0}|k, m \rangle = m |k, m \rangle . \label{Kzero}
\end{equation}
From the relations (\ref{hermi}), (\ref{Q}) and (\ref{Kmap}) it is
clear that the operators $\hat{K}_{\pm}$ act on the above states as
\begin{equation}
\hat{K}_{+}|k, m \rangle = \sqrt{\Phi (m(m+1)) - \Phi (k(k-1))}
|k, m + 1 \rangle , \label{Kplus}
\end{equation}
\begin{equation}
\hat{K}_{-}|k, m \rangle = \sqrt{\Phi (m(m-1)) - \Phi (k(k-1))}
|k, m - 1 \rangle . \label{Kminus}
\end{equation}

For the usual $su(1,1)$ case ($p=1$), we wish to take the basis states
$|k, m\rangle $ from those of the unitary irreducible representations of the
group $SU(1,1)$. As is well-known, the representations of $SU(1,1)$ are
classified into \cite{IKG,Wyb,BR}: ~(i) the positive discrete series
$D^{+}_{n}(k)$, ~(ii) the negative discrete series $D^{-}_{n}(k)$,
~(iii) the principle continuous series $C_{n}(m_{0}, k)$, and ~(iv) the
supplementary continuous series $E_{n}(m_{0}, k)$. As for the polynomial
$su(1,1)$, however, the corresponding group and its representations are
not available. In the present work, we are only interested in
constructing a set of coherent states for discrete dynamics of the
polynomial $su(1,1)$. Therefore, we examine whether the positive discrete
series $D_{n}^{+}(k)$ of $SU(1,1)$, for which
\begin{equation}
k \in {\bf R}^{+}, ~~~~~~m-k \in {\bf N}_{0}, \label{disrep}
\end{equation}
is compatible with $su_{\Phi }(1,1)$. Here we have used the notation
${\bf N}_{0}={\bf N}\cup \{0\}=\{0, 1, 2, 3, ...\}$.

The compact generator $\hat{K}_{0}$ has been chosen to be hermitian, so
that its eigenvalues $m$ are real. As the property (\ref{Kmap}) for
$r=1$ indicates, the operators $\hat{K}_{\pm}$ map eigenstates $|k,
m\rangle $ of $\hat{K}_{0}$ into $|k, m \pm 1\rangle $,
respectively. Hence the value of $m$ increases or decreases by integer
units as
\begin{equation}
m = m_{0} + n
\end{equation}
where $m_{0} \in {\bf R}$ and $n \in {\bf Z}$. The eigenvalue $\Phi
(k(k-1))$ of the Casimir operator $\hat{\bf K}^{2}$ must be real. In fact the
structure function has been assumed to be a real function increasing
with its argument greater than or equal to $-1/4$. Therefore $k$ must
satisfy the conditions,
\begin{equation}
k(k-1)\in {\bf R} ~~~~~\mbox{and}~~~~~ \left(k - \frac{1}{2}\right)^{2}
\geq 0,
\end{equation}
from which follows
\begin{equation}
k \in {\bf R}.
\end{equation}
Furthermore, (\ref{hermi}) yields
\begin{equation}
\langle k,m|\hat{K}_{\pm}^{\dagger}\,\hat{K}_{\pm}|k,m\rangle
=\langle k,m|\hat{K}_{\mp}\,\hat{K}_{\pm}|k,m\rangle  \geq
0,
\end{equation}
and (\ref{Q}) and (\ref{Kzero}) lead to
\begin{equation}
\Phi \left(m(m\pm 1)\right) - \Phi \left(k(k-1)\right) \geq 0.
\label{cond}
\end{equation}
As $\Phi (x)$ is an increasing function, the $SU(1,1)$
discrete series (\ref{disrep}) satisfies these conditions with
$m_{0}=k$. Thus we may choose as the basis $\{|k,m \rangle\}$ for
$su_{\Phi }(1,1)$
\begin{equation}
k \in {\bf R}^{+}, ~~~~~m=k + n ~~~(n \in {\bf N}_{0}).
\end{equation}

In the above analysis, we have not explicitly used the polynomial
condition (\ref{phi}) even though the structure function $\Phi (x)$ was
assumed to be expressible as a MacLaurin series of $x$.

In view of the basis chosen above, we realize that it is more convenient
to characterize the basis states by means of the integral number $n$
rather than $m=k+n$. Thus we let the orthonormalized set $\{|k, n\rangle
\}$ span the representation space with $k \in {\bf R}^{+}$ and $n \in
{\bf N}_{0}$. On this basis we rewrite (\ref{Kzero}), (\ref{Kplus}) and
(\ref{Kminus}) as
\begin{equation}
\hat{K}_{0}|k, n \rangle = (k+n) |k, n \rangle ,  \label{Kzero2}
\end{equation}
\begin{equation}
\hat{K}_{+}|k, n \rangle = \sqrt{\phi_{n+1}(k)}
|k, n + 1 \rangle ,   \label{Kplus2}
\end{equation}
\begin{equation}
\hat{K}_{-}|k, n \rangle = \sqrt{\phi_{n}(k)}
|k, n - 1 \rangle  \label{Kminus2}
\end{equation}
where we have introduced the short-hand notation,
\begin{equation}
\phi _{n}(k)=\Phi ((k+n)(k+n-1)) - \Phi (k(k-1)), \label{phin}
\end{equation}
which we shall call the structure factor for convenience.
From (\ref{Kminus2}) it is evident that
\begin{equation}
\hat{K}_{-}|k, 0 \rangle  = 0.  \label{fidu}
\end{equation}
Hence $|k, 0\rangle$ can be taken as the fiducial state.
Also from (\ref{Kplus2}) follows that
\begin{equation}
|k, n \rangle = \frac{1}{\sqrt{[\phi_{n}(k)]!}}(\hat{K}_{+})^{n}
|k, 0 \rangle .   \label{K-n}
\end{equation}
In the above we have used the generalized factorial notation signifying
\begin{equation}
[\phi _{n}(k)]! = \prod_{l=1}^{n}\phi _{l}(k), ~~~~~~~~[\phi _{0}(k)]!=1,
\end{equation}
which will also be used later for other sequences of
functions. Furthermore, for simplicity, we express $\phi _{n}(k)$ by
$\phi _{n}$.

\section{Coherent states for $su_{\Phi }(1,1)$}

Now we wish to construct generalized coherent states for $su_{\Phi
}(1,1)$ which accommodate those of the Barut-Girardello type and the
Perelomov type as special cases. By the Barut-Girardelo type (BG-type)
and the Perelomov type (P-type), we mean the coherent states for the
nonlinear $su(1,1)$ which are reducible to the Barut-Girardello
$SU(1,1)$ coherent states and the Perelomov $SU(1,1)$ coherent states in
the linear limit, respectively.

\subsection{Generalized coherent states}

First we introduce a generalized exponential function,
\begin{equation}
[e(\nu )]^{x}=\sum_{n=0}^{\infty }\frac{x^{n}}{[\nu _{n}]!}
\label{[e]}
\end{equation}
defined on a base sequence $\{\nu _{1}, \nu _{2}, \cdots, \nu _{n}\}$
with $\lim_{n\rightarrow \infty }|\nu _{n}| \neq 0$.
Then we consider a set of states constructed on the fiducial state
(\ref{fidu}) as
\begin{equation}
|k, \zeta \rangle = N_{\Phi }^{-1}(|\zeta |)\,
[e(\nu )]^{\zeta \hat{K}_{+}}\,|k, 0\rangle, \label{CS0}
\end{equation}
where $\zeta  \in {\bf C}$.
This is similar in form to the definition of the Perelomov $SU(1,1)$
coherent states (\ref{P}). However, we take this as a unified treatment
of the BG-type and the P-type. By the definition of the generalized
exponential function (\ref{[e]}) the state (\ref{CS0}) is expressed as
\begin{equation}
|k, \zeta \rangle = N_{\Phi }^{-1}(|\zeta |)
\sum_{n=0}^{\infty }\frac{(\zeta\hat{K}_{+})^{n}}{[\nu _{n}]!}
|k, 0\rangle . \label{CS01}
\end{equation}
Use of (\ref{K-n}) further leads (\ref{CS01}) to an alternative form,
\begin{equation}
|k, \zeta \rangle = N_{\Phi }^{-1}(|\zeta |)\sum_{n=0}^{\infty
}\frac{\sqrt{[\phi _{n}]!}}{[\nu _{n}]!}\,\zeta^{n}|k, n \rangle .
\label{CS02}
\end{equation}
These states are normalized to unity with
\begin{equation}
|N_{\Phi }(|\zeta |)|^{2}=\sum_{n=0}^{\infty } \,
\frac{[\phi_{n}]!}{([\nu_{n}]!)^{2}}|\zeta  |^{2n}. \label{Norm}
\end{equation}
Here the radius of convergence is
\begin{equation}
R = \lim_{n \rightarrow \infty } \frac{|\nu _{n}|^{2}}{|\phi _{n}|}.
\end{equation}

The states (\ref{CS02}), parameterized by a continuous complex number
$\zeta $, share a number of the properties that the coherent states are
to possess. They are not in general orthogonal. From the Schwarz
inequality, we have
\begin{equation}
\langle k, \zeta |k, \zeta '\rangle = N_{_{\Phi }}^{\ast \,-1}(|\zeta |) N_{_{\Phi
}}^{-1}(|\zeta'|)\,\sum_{n=0}^{\infty }\frac{[\phi _{n}]!}{([\nu
_{n}]!)^{2}} (\zeta ^{\ast}\zeta ')^{n}\leq 1, \label{ineq}
\end{equation}
which is not zero when $\zeta \neq \zeta '$. They resolve unity,
\begin{equation}
\hat{1}=\int d\mu (\zeta , \zeta ^{\ast}) \,|k, \zeta \rangle \langle k,
\zeta |,
\label{unity}
\end{equation}
if the integration measure can be found in the form,
\begin{equation}
d\mu (\zeta , \zeta ^{\ast})=\frac{1}{2\pi }\,|N_{\Phi }(|\zeta
|)|^{2}\,
\rho (|\zeta |^{2})\,d|\zeta |^{2}\,d\varphi . \label{meas}
\end{equation}
Here $\zeta =|\zeta |\,e^{i\varphi } ~(0 \leq \varphi < 2\pi )$, and the weight
function $\rho (|\zeta |^{2})$ is to be determined by its
moments,
\begin{equation}
\int_{0}^{\infty } \,\rho (t)\, t^{n}\, dt=\frac{([\nu
_{n}]!)^{2}}{[\phi _{n}]!}, \label{weight}
\end{equation}
where we have let $t=|\zeta |^{2}$. The non-orthogonality (\ref{ineq})
together with the resolution of unity (\ref{unity}) show that the states
form an overcomplete basis in the representation space spanned by the
discrete eigenstates of the compact operator $\hat{K}_{0}$ bounded
below. Note also that these states are temporally stable for a system
with the Hamiltonian $\hat{H}=\hbar \omega (\hat{K}_{0}-k)$ as the states
(\ref{CS02}) evolve according to
\begin{equation}
e^{-i\hat{H}t/\hbar}|k, \zeta \rangle = |k, \zeta \,e^{-i\omega t} \rangle .
\end{equation}
With these properties the states constructed in (\ref{CS02})
may be considered as generalized coherent states for $su_{\Phi }(1,1)$.

\subsection{Coherent states for $su_{2p-1}(1,1)$}

Next we impose on $su_{\Phi }(1,1)$ the polynomial condition,
\begin{equation}
\Phi (x)=\sum_{r=1}^{p}\alpha _{r}\,x^{r} ~~~~~(\alpha _{r} \in {\bf R})
\label{phi2}
\end{equation}
where $\alpha _{1} > 0$,  $\alpha _{p} \neq 0$, $d\Phi /dx > 0$ and $x
\geq -1/4$. This is the same as (\ref{phi}) applied to $su_{\Phi
}(2)$. Under this condition, $su_{\Phi }(1,1)$ becomes a polynomial
$su(1,1)$ algebra of order $2p-1$, which we denote by $su_{2p-1}(1,1)$.
In the limit that $\alpha _{r} \rightarrow 0$ for $r=2,3,...,p$, the
structure function for $p=1$ becomes $\Phi (x)=\alpha _{1}x$. In the
resultant linear algebra $su_{1}(1,1)$, we can let $\alpha _{1}=1$
without loss of generality. Thus we identify $su_{1}(1,1)$ with the
usual linear $su(1,1)$ algebra.

For $su_{2p-1}(1,1)$ the structure factor $\phi _{n}$ of (\ref{phin})
takes the form
\begin{equation}
\phi _{n}= \sum_{r=1}^{p}\alpha _{r}\left[(k+n)^{r}(k+n-1)^{r} -
k^{r}(k-1)^{r}\right]= n\,(2k + n -1)\,\chi _{n} \label{phichi}
\end{equation}
where
\begin{equation}
\chi _{n}= \sum_{r=1}^{p}\sum_{s=1}^{r}\,\alpha _{r}[k(k-1)]^{r-
s}[(k+n)(k+n-1)]^{s-1}. \label{chi}
\end{equation}
Note that for large $n$
\begin{equation}
\chi _{n} \sim O(n^{2p-2}), ~~~~~\phi_{n} \sim  O(n^{2p}). \label{radius}
\end{equation}
It is evident that $\chi _{n} = \alpha _{1}$ and $\phi _{n} = n(2k+n-1)$
for $p=1$. This means that $\chi _{n}$ for $p > 1$ characterizes the
nonlinear deformation of $su_{2p-1}(1,1)$. In this regard, we refer to
$\chi _{n}$ as the deformation factor.

The generalized factorial of $\phi_{n}$ given by (\ref{phichi}) is
\begin{equation}
[\phi _{n}]! =n!\,(2k)_{n}\,[\chi _{n}]!,  \label{phif}
\end{equation}
where used is the Pochhammer symbol,
\begin{equation}
(z)_{n}=\frac{\Gamma (z+n)}{\Gamma (z)}=(-1)^{n}\frac{\Gamma (1-
z)}{\Gamma(1-z-n)}. \label{symb}
\end{equation}

The deformation factor $\chi _{n}$ of (\ref{chi}) is an
inhomogeneous polynomial of degree $2p-2$ which can be written as
\begin{equation}
\chi _{n}= \alpha _{p}\,\prod_{i=1}^{2p-2}(n-a_{i}) \label{chip}
\end{equation}
where $a_{i}$'s are the roots of $\chi _{n} = 0$ with respect to $n$.
Its generalized factorial can be expressed as
\begin{equation}
[\chi _{n}]! = \chi _{1}\chi _{2} \dots \chi _{n} = \alpha
_{p}^{n}\,\prod_{i=1}^{2p-2}\,(1-a_{i})_{n}\,. \label{chif}
\end{equation}
Substitution of (\ref{chif}) into (\ref{phif}) yields
\begin{equation}
[\phi _{n}]!=\alpha _{p}^{n}\,n!\,(2k)_{n}\,\prod_{i=1}^{2p-2}(1-
a_{i})_{n}
\label{phin2}
\end{equation}

Inserting (\ref{phif}) into (\ref{CS02}) and (\ref{Norm}), we obtain a formal
expression for the coherent states for the polynomial algebra
$su_{2p-1}(1,1)$,
\begin{equation}
|k, \zeta  \rangle = N_{p}^{-1}(|\xi |)\,\sum_{n=0}^{\infty }
\frac{\sqrt{n!\,(2k)_{n}\,[\chi _{n}]!}}{[\nu _{n}]!}
\zeta ^{n}|k, n \rangle ,  \label{CS03}
\end{equation}
and
\begin{equation}
|N_{p}(|\zeta |)|^{2} = \sum_{n=0}^{\infty }\,
\frac{n!\,(2k)_{n}\,[\chi _{n}]!}{([\nu _{n}]!)^{2}}\,
|\zeta |^{2n}. \label{Np}
\end{equation}
The coherent states (\ref{CS03}) remain to be formal until $[\nu
_{n}]!$ is specified. In order to accommodate the set of $SU(1,1)$
coherent states as a limiting case, we have to choose appropriately
$[\nu _{n}]!$. In the proceeding sections we specifically consider
two cases: the Barut-Girardello type (BG-type) whose states go over to
the Barut-Girardello $SU(1,1)$ states in the linear limit ($p=1$), and
the Perelomov type (P-type) whose coherent states approach the Perelomov
$SU(1,1)$ states in the same limit.

\section{Coherent states of the Barut-Girardello type}

Out of the generalized coherent states (\ref{CS03}) formally constructed
for the $su_{2p-1}(1,1)$ we select the BG-type states by letting
\begin{equation}
\nu _{n}= \phi _{n}.
\end{equation}
With this choice, (\ref{CS02}) reads
\begin{equation}
|k, \xi \rangle = N_{p}^{-1}(|\xi |)\,
\sum_{n=0}^{\infty }\,\frac{\xi
^{n}}{\sqrt{[\phi _{n}]!}}\,|k, n\rangle \label{BG01}
\end{equation}
where we have let $\zeta = \xi $ for the BG-type. Because of
(\ref{radius}), the radius of convergence of (\ref{BG01}) is infinity.
This means that the BG-type states (\ref{BG01}) can be defined on the
full complex plane of $\xi $.
It is easy to verify by utilizing (\ref{Kplus2}) that the coherent
states (\ref{BG01}) are indeed eigenstates of the non-hermitian operator
$\hat{K}_{+}$,
\begin{equation}
\hat{K}_{+}|k, \xi \rangle = \xi \,|k, \xi \rangle
\end{equation}
with complex eigenvalues $\xi $. More explicitly, substitution of
(\ref{phin2}) into (\ref{BG01}) yields
\begin{equation}
|k, \xi \rangle = N_{p}^{-1}(|\xi |)\,\sum_{n=0}^{\infty }\,
\left\{
\alpha _{p}^{n}\,n!\,(2k)_{n}\,\prod_{i=1}^{2p-2}(1-a_{i})_{n}\right\}^{-1/2}\,\xi^{n}
|k, n\rangle. \label{BG02}
\end{equation}
The normalization factor is
\begin{equation}
|N_{p}(|\xi |)|^{2}=\sum_{n=0}^{\infty } \,\frac{1}{n!\,(2k)_{n}\,
\prod_{i=1}^{2p-2}(1-a_{i})_{n}} \left(\frac{|\xi |^{2}}{\alpha
_{p}}\right)^{n}
\end{equation}
which can be expressed in closed form as
\begin{equation}
|N_{p}(|\xi |)|^{2}=\,_{0}F_{2p-1}\left(
2k, 1-a_{1}, 1-a_{2}, \cdots , 1-a_{2p-2}; |\xi |^{2}/\alpha
_{p}\right), \label{normp}
\end{equation}
where $ \,_{p}F_{q}$ is Pochhammer's generalized hypergeometric function
defined by
\[
\,_{p}F_{q}(\alpha _{1}, \alpha _{2}, \cdots \alpha _{p}; \gamma _{1},
\gamma _{2}, \cdots \gamma_{q}; z)=\sum_{n=0}^{\infty }
\frac{(\alpha _{1})_{n}(\alpha _{2})_{n}\cdots (\alpha _{p})_{n}}{
(\gamma _{1})_{n}(\gamma_{2})_{n}\cdots
(\gamma_{q})_{n}}\frac{z^{n}}{n!} .
\]
The hypergeometric series $\,_{0}F_{q}$ is analytic at any $z$. Hence
the normalization factor (\ref{normp}) is convergent for all values of
$|\xi |^{2}/\alpha _{p}$.

The inner product of two such states takes the form,
\begin{equation}
\langle k, \xi |k, \xi' \rangle = N_{p}^{\ast \,-1}(|\xi |)N_{p}^{-
1}(|\xi '|)
\,_{0}F_{2p-1}\,\left(2k, 1-a_{1}, 1-a_{2}, \cdots , 1-
a_{2p-2}; \,\xi ^{\ast}\xi '/\alpha _{p}\right).
\end{equation}

The coherent states thus constructed for $su_{2p-1}(1,1)$ are able to
resolve unity if the weight function $\rho (|\xi |^{2})$ is determined
as follows. Inserting (\ref{phin2}) into (\ref{weight}) we obtain
\begin{equation}
\int_{0}^{\infty } \rho (t)\,t^{n}\,dt =
\alpha _{p}^{n}\,n!\,(2k)_{n}\,\prod_{i=1}^{2p-2}(1-
a_{i})_{n}
\end{equation}
or rewriting with $n=s-1$
\begin{equation}
\int_{0}^{\infty } \rho (t)\,t^{s-1}\,dt = (\alpha _{p})^{s-
1}\frac{\Gamma (s)\,\Gamma (2k-1+s)\,\Gamma (-a_{1} + s)\,\Gamma (-
a_{2}+s) \cdots \Gamma (-a_{2p-2}+s)}{ \Gamma (2k)\,\Gamma (1-
a_{1})\,\Gamma (1-a_{2}) \cdots \Gamma (1-a_{2p-2})},
\end{equation}
from which the weight function can be found by the inverse Mellin
transformation (see Formula 7.811.4 in \cite{GR}) in terms of Meijer's
$G$-function as
\begin{equation}
\rho (|\xi |^{2})=\left[\alpha _{p}\,\Gamma (2k) \prod_{i=1}^{2p-
2}\Gamma (1-a_{i})\right]^{-1} G_{0\, 2p}^{2p\, 0}\left(\frac{|\xi
|^{2}}{\alpha _{p}} \left|\begin{array}{c}~\\~\end{array} 0, \,2k-1, \,-
a_{1}, \,-a_{2}, \cdots , \,-a_{2p-2} \right. \right). \label{weight2}
\end{equation}
With the weight function (\ref{weight2}) for the measure
(\ref{meas}) the resolution of unity (\ref{unity}) can be achieved.

So far we have selected the BG-type coherent states (\ref{BG02}) out of
the generalized coherent states (\ref{CS02}). It is rather
straightforward to show that the constructed states (\ref{BG02}) are
indeed reducible to the Barut-Girardello $SU(1,1)$ states in the linear
limit. If the deformation factor tends to unity, i.e., $\chi _{n}
\rightarrow 1$, then $[\phi _{n}]! \rightarrow n!\,(2k)_{n}$. For $p=1$
and $\alpha _{1}=1$, the normalization factor (\ref{normp}) takes the
form,
\begin{equation}
|N_{1}(|\xi |^{2})|^{2} =\,_{0}F_{1}(2k; |\xi |^{2})
=\Gamma (2k)|\xi |^{1-2k}\,I_{2k-1}(2|\xi |),
\end{equation}
where $I_{\nu }(z)$ is the modified Bessel function of the first kind.
Thus in the linear limit the coherent states (\ref{BG02})
becomes
\begin{equation}
|k, \xi \rangle = N_{1}^{-1}(|\xi |)\sum_{n=0}^{\infty }
\frac{1}{\sqrt{n!\,(2k)_{n}}} \, \xi ^{n}|k, n \rangle.  \label{BG0}
\end{equation}
The coherent states (\ref{BG0}) are indeed the Barut-Girardello $SU(1,1)$
coherent states \cite{BG}. The weight function that enables the states
(\ref{BG0}) to resolve the unity follows from
\begin{equation}
\int_{0}^{\infty } \rho (t)\,t^{s-1}\,dt = \Gamma (s)\,(2k)_{s-1},
\end{equation}
the result being
\begin{equation}
\rho (|\xi |^{2})=\frac{1}{\Gamma (2k)}\,
G_{0 \,2}^{2\, 0}\left(|\xi |^{2} \left| 0, 2k-1 \right. \right)
=\frac{2|\xi |^{2k-1}}{\Gamma (2k)}\,K_{2k-1}(2|\xi |)
\end{equation}
where $K_{\nu }(z)$ is the modified Bessel function of the second kind.

\section{Coherent states of the Perelomov type}

Our next task is to construct a set of the Perelomov type states from
(\ref{CS03}). To this end, we choose
\begin{equation}
\nu _{n}= n\,\chi _{n}
\end{equation}
and let $\zeta =\eta $ to write (\ref{CS03}) in the form,
\begin{equation}
|k, \eta \rangle = N_{p}^{-1}(|\eta |)\,\sum_{n=0}^{\infty
}\frac{\sqrt{[\phi _{n}]!}}{n!\,[\chi _{n}]!}\,\eta ^{n}|k, n\rangle,
\label{CSPp}
\end{equation}
or, using (\ref{phif}),
\begin{equation}
|k, \eta \rangle = N_{p}^{-1}(|\eta |)\,\sum_{n=0}^{\infty
}\sqrt{\frac{(2k)_{n}}{n!\,[\chi _{n}]!}}\,\eta ^{n}|k, n\rangle.
\label{CSPp2}
\end{equation}
The radius of convergence for (\ref{CSPp2}) is obtained by
\begin{equation}
R = \lim_{n \rightarrow \infty }\frac{|n\chi _{n}|^{2}}{n(2k+n)|\chi
_{n}|}=\lim_{n \rightarrow \infty }|\chi _{n}|,
\end{equation}
whose result depends on the parameter $p$. Since $\chi _{n}=\alpha _{1}$
for $p=1$, the radius of convergence is finite, i.e., $R=\alpha _{1}$.
If $p \neq 1$, again from (\ref{radius}), the radius $R$ becomes
infinity. Substitution of (\ref{phin2}) and (\ref{chip}) converts
(\ref{CSPp2}) into
\begin{equation}
|k, \eta \rangle = N_{p}^{-1}(|\eta |)\,\sum_{n=0}^{\infty
}\left[\frac{(2k)_{n}}{n!\,\prod_{j=1}^{2p-2}(1-a_{j})_{n}}
\right]^{1/2}\,\left(\frac{\eta }{\sqrt{\alpha _{p}}}\right)^{n}|k, n\rangle, \label{CSP2}
\end{equation}
where the normalization factor in (\ref{CSP2}) is given by
\begin{equation}
|N_{p}(|\eta |)|^{2}= \,_{1}F_{2p-2}\left(2k; 1-a_{1}, 1-a_{2},
\cdots, 1-a_{2p-2}; |\eta |^{2}/\alpha _{p}\right) \label{Normp}
\end{equation}
which is convergent for any real value of $|\eta |^{2}/\alpha _{p}$ if
$p > 1$. The weight function $\rho (|\eta |^{2})$ needed to resolve the
unity can be determined by
\begin{equation}
\int_{0}^{\infty }\rho (t)\,t^{n}\,dt = \frac{n!\,[\chi
_{n}]!}{(2k)_{n}}.
\end{equation}
Utilizing $[\chi _{n}]!$ of (\ref{chif}) and letting $n=s-1$, we rewrite
this as
\begin{equation}
\int_{0}^{\infty }\rho (t)\,t^{s-1}\,dt = \frac{\Gamma (2k)}
{\prod_{j=1}^{2p-2}\Gamma (1-a_{j})}\, \frac{\alpha _{p}^{s-1}\Gamma
(s)}{\Gamma (2k-1 +s)}\,\prod_{j=1}^{2p-2}\Gamma (-a_{j} + s)
\end{equation}
from which we obtain the weight function
\begin{equation}
\rho (|\eta |^{2})=\frac{\Gamma (2k)}{\alpha _{p}\,\prod_{j=1}^{2p-2}
\Gamma (1-a_{j})}\,
G_{1\, ~2p-1}^{2p-1\, ~0}\left(\frac{|\eta |^{2}}{\alpha _{p}}
~\left|\begin{array}{c}2k-1 \\ 0, -a_{1}, -a_{2}, \cdots , -a_{2p-2}
\end{array}~\right. \right) \label{rhoc}
\end{equation}
valid for all values of $|\eta |^{2}/\alpha _{p}$ if $p > 1$, and for $
0 < |\eta |^{2}/\alpha _{1} < 1$ if $p=1$.

In the linear limit $\chi _{n} \rightarrow \alpha _{1}$, the
normalization factor (\ref{Normp}) tends to
\begin{equation}
|N_{1}(|\eta |)|^{2}= \,_{1}F_{0}\left(2k; |\eta |^{2}/\alpha
_{1}\right) = (1 - |\eta |^{2}/\alpha _{1})^{-2k}.
\end{equation}
Therefore the coherent states (\ref{CSP2}) becomes
\begin{equation}
|k, \eta \rangle = (1 - |\eta |^{2}/\alpha _{1})^{k}\,\sum_{n=0}^{\infty
}\left[\frac{(2k)_{n}}{n!}
\right]^{1/2}\,\left(\frac{\eta }{\sqrt{\alpha _{1}}}\right)^{n}|k, n\rangle.
\label{CSPs}
\end{equation}
With $\alpha _{1}=1$, the last expression (\ref{CSPs}) coincides with
Perelomov's result for the $SU(1,1)$ coherent states \cite{Pere}.
For $p=1$ and $\alpha _{1}=1$ the weight function (\ref{rhoc}) reduces
to
\begin{equation}
\rho (|\eta |^{2})=\Gamma (2k)
G_{1\, ~1}^{1\, ~0}\left(|\eta |^{2}
~\left|\begin{array}{c}2k-1\\ 0\end{array} \right. \right). \label{rhos}
\end{equation}
With the help of the identity
\begin{equation}
G_{1\, ~1}^{1\, ~0}\left(z ~\left|\begin{array}{c}2k-1\\ 0 \end{array}
\right. \right) = \frac{1}{\Gamma (2k-1)} \,_{1}F_{0}(2-2k; z)
=\frac{1}{\Gamma (2k-1)} \,(1-z)^{2k-2},
\end{equation}
valid for $0 < |z| <1$, the weight function can be simplified to the form
\begin{equation}
\rho (|\eta |^{2}) = (2k-1)\left(1 - |\eta |^{2}\right)^{2k-2}
\end{equation}
which is defined only on the Poincar\'e disk. Furthermore, in order for
the weight function to remain positive, it is necessary to demand that
$2k > 1$.

\section{Coherent states for the cubic algebra}

In this section, we study the cubic case in more detail with one
of the conditionally solvable problems in supersymmetric (SUSY) quantum
mechanics proposed by Junker and Roy \cite{JR}.

\subsection{The cubic $su(1,1)$ algebra}

The cubic $su(1,1)$ algebra ($p=2$) is the simplest special case of the
odd-polynomial $su_{2p-1}(1,1)$ for which the structure function is
quadratic,
\begin{equation}
\Phi (x) = \alpha _{1} x + \alpha _{2}x^{2}, \label{quad}
\end{equation}
where $\alpha _{1} > 0$ and $\alpha _{2}\neq 0$. The deformed algebra
(\ref{sup11}) with this quadratic structure function becomes a cubic
algebra of the form,
\begin{equation}
[\hat{K}_{0}, \hat{K}_{\pm}]=\pm \hat{K}_{\pm}, ~~~~~
[\hat{K}_{+}, \hat{K}_{-}] = -2\alpha _{1}
\hat{K}_{0} - 4\alpha _{2} \hat{K}_{0}^{3}.  \label{3su11}
\end{equation}
The deformation factor for the cubic algebra is
\begin{equation}
\chi _{n}= \alpha _{1} + \alpha _{2}\{(k+n)(k+n-1)+k(k-1)\}
\label{chi21}
\end{equation}
which can be written as
\begin{equation}
\chi _{n} = \alpha _{2}(n-a_{+})(n-a_{-})  \label{chi2}
\end{equation}
with the roots
\begin{equation}
a_{\pm} =- \frac{1}{2}(2k-1) \pm \frac{1}{2}\left\{2 - (2k-1)^{2}
- 4\frac{\alpha _{1}}{\alpha _{2}}\right\}^{1/2}. \label{roots}
\end{equation}
Hence the structure factor defined by (\ref{phichi}) reads
\begin{equation}
\phi _{n}(k) = \alpha _{2}n (2k + n -1)(n - a_{+}) (n - a_{-}),
\label{phi3}
\end{equation}
with which the ladder operators $\hat{K}_{+}$ and $\hat{K}_{-}$ work
in the representation space of $su_{3}(1,1)$ as
\begin{equation}
\hat{K}_{+}|k, n\rangle =
\sqrt{\alpha _{2}(n+1)(2k + n)(n+1 - a_{+}) (n+1 - a_{-})} \,|k, n + 1\rangle
\end{equation}
\begin{equation}
\hat{K}_{-}|k, n\rangle =
\sqrt{\alpha _{2}n (2k + n -1)(n - a_{+}) (n - a_{-})} \,|k, n - 1\rangle.
\end{equation}
The coherent states of the BG-type and the P-type can be constructed
straightforwardly for the cubic algebra.

\subsection{Conditionally solvable problems}

At this point we reformulate the conditionally solvable broken SUSY problem in
\cite{JR} in a way appropriate to the present polynomial $su(1,1)$
scheme.

In SUSY quantum mechanics (see, e.g., \cite{Junk}), the partner
Hamiltonians are given by
\begin{equation}
\hat{H}_{\pm}=\frac{1}{2}\hat{p}^{2} + V_{\pm}(\hat{x}). \label{Hamil}
\end{equation}
The partner potentials are expressed in terms of the SUSY potential
$W(x)$ as
\begin{equation}
V_{\pm}(x)=\frac{1}{2}\left\{W^{2}(\hat{x}) \pm i[\hat{p}, W(\hat{x})]
\right\} \label{Vpm}
\end{equation}
where $[\hat{x}, \hat{p}]=i ~(\hbar=1)$.
The partner Hamiltonians (\ref{Hamil}) may also be written as
\begin{equation}
\hat{H}_{+} = \hat{A}\hat{A}^{\dagger}, ~~~~~~~~
H_{-} = \hat{A}^{\dagger}\hat{A},
\end{equation}
where
\begin{equation}
\hat{A}=\frac{1}{\sqrt{2}}\,\left(i\hat{p} + W(\hat{x})\right), ~~~~~~~~
\hat{A}^{\dagger}=\frac{1}{\sqrt{2}}\,\left(-i\hat{p} +
W(\hat{x})\right).
\end{equation}

Let the partner eigenequations be expressed by
\begin{equation}
\hat{H}_{\pm}|\psi _{n}^{(\pm)}\rangle =
E_{n}^{(\pm )}|\psi _{n}^{(\pm)}\rangle, ~~~~~n=0, 1, 2,...
\end{equation}
If SUSY is broken \cite{Junk},
\begin{equation}
E_{n}^{(+)} = E_{n}^{(-)} > 0,
\end{equation}
and
\begin{equation}
\hat{A}^{\dagger}|\psi _{n}^{(+)}\rangle=
\sqrt{E_{n}^{(+)}}|\psi _{n}^{(-)}\rangle ,
~~~~~~\hat{A}|\psi _{n}^{(-)}\rangle=
\sqrt{E_{n}^{(-)}}|\psi _{n}^{(+)}\rangle.  \label{psi}
\end{equation}

By definition, for conditionally solvable problems \cite{JR}, the
SUSY potential $W(\hat{x})$ is separable to two parts as
\begin{equation}
W(\hat{x})=U(\hat{x}) + f(\hat{x})
\end{equation}
where $U(x)$ is a shape-invariant SUSY potential and $f(x)$ is a function
satisfying the equation,
\begin{equation}
f^{2}(\hat{x}) + 2U(\hat{x})\,f(\hat{x}) + i[\hat{p}, f(\hat{x})] = 2
(\varepsilon -1),
\end{equation}
$\varepsilon $ being the adjustable parameter a certain value of which
makes the problem solvable. The partner potentials are written as
\begin{equation}
V_{+}(\hat{x})=\frac{1}{2}\left(U^{2}(\hat{x}) + i[\hat{p},
U(\hat{x})]\right) + \varepsilon - 1,
\end{equation}
\begin{equation}
V_{-}(\hat{x})=\frac{1}{2}\left(U^{2}(\hat{x}) - i[\hat{p},
U(\hat{x})]\right) - i[\hat{p}, f(\hat{x})] + \varepsilon - 1.
\end{equation}
Since $V_{+}(\hat{x})$ is a shape-invariant potential, the system of
$\hat{H}_{+}$ is exactly solvable. The potential $V_{-}(\hat{x})$ is not
shape-invariant, but the eigenvalue problem with $\hat{H}_{-}$ becomes
conditionally solvable.

As a specific example, we take, as in \cite{JR}, a modified radial harmonic
oscillator with broken SUSY, for which
\begin{equation}
U(x) = x + \frac{\gamma +1}{x} ~~~~(\gamma \geq 0), \label{spot}
\end{equation}
and
\begin{equation}
f(x) = \frac{d}{dx}\ln \,_{1}F_{1}\left(\frac{1}{2}-\frac{\varepsilon
}{2}, \gamma + \frac{3}{2}; -x^{2}\right)
\end{equation}
in the coordinate representation. In order for the confluent
hypergeometric function to be convergent for the whole range of $x$, the
parameter $\varepsilon $ must be subjected to the condition,
\begin{equation}
\varepsilon + 2\varepsilon\gamma  + 2 >0. \label{epsilon}
\end{equation}
This is indeed the condition on $\varepsilon $ under which the modified
oscillator becomes exactly solvable.

The potential $V_{+}(x)$ composed of the SUSY potential (\ref{spot}) is
\begin{equation}
V_{+}(x) =\frac{1}{2}x^{2} + \frac{\gamma (\gamma +1)}{x^{2}} + \gamma +
\varepsilon + \frac{1}{2},
\end{equation}
which is shape-invariant by choice. Although the exact energy spectrum
of the Hamiltonian $\hat{H}_{+}$ can be calculated by the standard
Gendenstein procedure \cite{Gen} or by using the semiclassical broken
SUSY formula \cite{IJ}, we employ here an algebraic approach
\cite{IKG,BR}. To this end we introduce the following operators,
\begin{equation}\begin{array}{ll}
\hat{C}_{0}&=\frac{1}{2}\left(\hat{H}_{+} - g\hat{1}\right)\\
\hat{C}_{1}&=\frac{1}{4}\left(\hat{p}^{2} - \hat{x}^{2} + \frac{\gamma
(\gamma +1)}{\hat{x}^{2}}\right)\\
\hat{C}_{2}&=\frac{1}{4}\left(\hat{x}\hat{p} + \hat{p}\hat{x} \right)
\end{array}
\end{equation}
where $g=\gamma + \varepsilon + 1/2$. It is then easy to show that they
obey the $su(1,1)$ algebra,
\begin{equation}
[\hat{C}_{0}, \hat{C}_{\pm}] = \pm \hat{C}_{\pm}, ~~~~~~
[\hat{C}_{+}, \hat{C}_{-}] = -2 \hat{C}_{0}.   \label{su11}
\end{equation}
where $\hat{C}_{\pm} \equiv \hat{C}_{1} \pm i \hat{C}_{2}$. The Casimir
operator is
\begin{equation}
\hat{\bf C}^{2} \equiv  \hat{C}_{0}^{2} - \hat{C}_{1}^{2} - \hat{C}_{2}^{2}
\label{CasiA1}
\end{equation}
which turns out to be
\begin{equation}
\hat{\bf C}^{2} = \frac{4\gamma (\gamma +1)-3}{16}\hat{1}. \label{C-Casi}
\end{equation}
On the basis $\{|c, n\rangle \}$ that diagonalizes $\hat{\bf C}^{2}$ and
$\hat{C}_{0}$ simultaneously,
\begin{equation}
{\bf C}^{2}|c, n\rangle = c(c-1)|c, n\rangle , ~~~~~~
C_{0}|c, n\rangle = (c+n)|c, n\rangle ,      \label{C-C}
\end{equation}
where $c \in {\bf R}^{+}$ and $n \in {\bf N}_{0}$. From (\ref{C-Casi})
and (\ref{C-C}), we recognize that the modified radial harmonic
oscillator under consideration is characterized by the constant,
\begin{equation}
c= \frac{1}{4}\left(2\gamma + 3\right), \label{c}
\end{equation}
and that the spectrum of $\hat{H}_{+}$ is
\begin{equation}
E_{n}^{(+)}=2(c+n) + g = 2n + 2\gamma + 2 + \varepsilon. \label{Eplus}
\end{equation}
Since the Hamiltonian $\hat{H}_{+}$ is diagonalized on the basis that
diagonalizes the operator $\hat{C}_{0}$, we identify the $su(1,1)$
states $|c, n\rangle $ characterized by (\ref{c}) with the eigenstates
$|\psi _{n}^{+}\rangle $ of $\hat{H}_{+}$.
Thus the ladder operators act on the SUSY states as
\begin{equation}
\hat{C}_{+}|\psi _{n}^{(+)}\rangle =
\sqrt{(n+1)(n + \gamma + 3/2)}\,|\psi _{n+1}^{(+)}\rangle ,
\label{laddCp}
\end{equation}
\begin{equation}
\hat{C}_{-}|\psi _{n}^{(+)}\rangle= \sqrt{n(n + \gamma + 1/2)} \,|
\psi _{n-1}^{(+)}\rangle . \label{laddCm}
\end{equation}

Next we define the operators
\begin{equation}
\hat{D}_{0}=\frac{1}{2}\hat{H}_{-}=\frac{1}{2}\hat{A}^{\dagger}\hat{A},
~~~~~~\hat{D}_{\pm}=\hat{A}^{\dagger}\hat{C}_{\pm}\hat{A}. \label{Ds}
\end{equation}
Use of (\ref{psi}), (\ref{laddCp}), and (\ref{laddCm}) enables us to
show that $\hat{D}_{\pm}$, when acting on the SUSY states
$|\psi _{n}^{(-)}\rangle$, behave like the ladder operators,
\begin{equation}
\hat{D}_{+}|\psi _{n}^{(-)}\rangle = \sqrt{E_{n}^{(-)}} \sqrt{(n+1)(n +
\gamma + 3/2)}\sqrt{E_{n+1}^{(+)}} \,|\psi _{n+1}^{(-)}\rangle ,
\label{rai1}
\end{equation}
and
\begin{equation}
\hat{D}_{-}|\psi _{n}^{(-)}\rangle = \sqrt{E_{n}^{(-)}} \sqrt{n(n +
\gamma + 1/2)}\sqrt{E_{n-1}^{(+)}} \,|\psi _{n-1}^{(-)}\rangle .
\label{low1}
\end{equation}
What we wish to stress here is that the operators introduced by
(\ref{Ds}) form a cubic algebra,
\begin{equation}
[\hat{D}_{0}, \hat{D}_{\pm}] = \pm \hat{D}_{\pm}, ~~~~~
[\hat{D}_{+}, \hat{D}_{-}] = -2\left(g^{2} - (2c
-1)^{2} +1\right)\hat{D}_{0} + 12 g \hat{D}_{0}^{2} - 16\hat{D}_{0}^{3},
\label{Dalgeb}
\end{equation}
where $g=\gamma +\varepsilon +1/2$ and $c=(2\gamma +3)/4$. This algebra
contains a quadratic term. It is not certain whether the representation
we have constructed for the odd-polynomial algebra $su_{_{2p-1}}(1,1)$
in section 2 is applicable to this case. Therefore we select the
parameter $\varepsilon $ such that $g=0$. Then we have the
odd-polynomial cubic $su(1,1)$ algebra of interest,
\begin{equation}
[\hat{D}_{0}, \hat{D}_{\pm}] = \pm \hat{D}_{\pm}, ~~~~~
[\hat{D}_{+}, \hat{D}_{-}] = -\left(\frac{3}{2} - 2\gamma (\gamma + 1)
\right)\hat{D}_{0} - 16\hat{D}_{0}^{3}
\label{Dcub}
\end{equation}
provided that
\begin{equation}
3 -4\gamma (\gamma +1) > 0. \label{g0}
\end{equation}
The two conditions (\ref{epsilon}) and (\ref{g0}) lead us to the
restrictions on $\varepsilon $ or $\gamma $,
\begin{equation}
-1 < \varepsilon < \frac{1}{2} ~~~~~\mbox{or}~~~~~ 0 < \gamma  <
\frac{1}{2},
\end{equation}
under which we shall work now on.

By comparing (\ref{Dcub}) with the cubic algebra (\ref{3su11}), we
determine the parameters of (\ref{quad})
\begin{equation}
\alpha _{1} = \frac{3}{4} - \gamma (\gamma + 1)
 ~~~~~~~~~\alpha _{2}=4, \label{alpha}
\end{equation}
from which follows the structure function,
\begin{equation}
\Phi (x)=\left\{\frac{3}{4} - \gamma (\gamma + 1)\right\}x + 4x^{2}.
\label{3Phi}
\end{equation}
From (\ref{Q}) the Casimir operator for the cubic algebra (\ref{Dcub})
is given by
\begin{equation}
{\bf D}^{2}=-\hat{D}_{+}\hat{D}_{-} + \left\{\frac{3}{4} - \gamma
(\gamma + 1)\right\}\hat{D}_{0}(\hat{D}_{0}+1) +
4\hat{D}_{0}^{2}(\hat{D}_{0} + 1)^{2}.
\end{equation}
With the basis $\{|d, n\rangle \}$, we diagonalize $\hat{D}_{0}$
in (\ref{Dcub}) and the Casimir operator
${\bf D}^{2}$ of the cubic algebra as
\begin{equation}
{\bf D}^{2}|d, n\rangle = d(d-1)|d, n\rangle , ~~~~~~
\hat{D}_{0}|d, n\rangle = (d + n)|d, n\rangle ,
\end{equation}
where $d \in {\bf R}^{+}$ and $n \in {\bf N}_{0}$. Since the operator
$\hat{H}_{-}$ is also diagonalized, we consider the $su_{_{3}}(1,1)$
states $|d, n\rangle $ as the eigenstates of $\hat{H}_{-}$ yielding the
spectrum,
\begin{equation}
E_{n}^{(-)}=2n + 2d. \label{Eminus}
\end{equation}
In broken SUSY, as is mentioned above, the spectra of the partner
Hamiltonians are identical, that is, $E_{n}^{(+)}=E_{n}^{(-)}=E_{n}$.
Hence, comparing (\ref{Eplus}) and (\ref{Eminus}) with the condition
$g=0$, we have
\begin{equation}
E_{n}= 2n + \gamma + \frac{3}{2}, ~~~~~(n=0, 1, 2, ...)
\end{equation}
This implies that the representation space of $su_{3}(1,1)$ is characterized
by the constant
\begin{equation}
d = \frac{1}{4}\left(2\gamma + 3\right).
\end{equation}
In this regard, we may identify the base states $|d, n\rangle$ of the
cubic algebra (\ref{Dcub}) with the eigenstates $|\psi _{n}^{(-)}\rangle
$ of $\hat{H}_{-}$. Even though the
characteristic constant $d$ of the representation of the cubic algebra
(\ref{Dcub}) coincides with the characteristic constant $c$, given by
(\ref{c}), of the $su(1,1)$ algebra (\ref{su11}), the two states $|c,
n\rangle$ and $|d, n\rangle$ are distinct; namely, as we have identified
in the above,
\begin{equation}
|\psi _{n}^{(+)}\rangle = |c, n\rangle, ~~~~~~~~~
|\psi _{n}^{(-)}\rangle = |d, n\rangle
\end{equation}
which are related by (\ref{psi}).

Substitution of the values (\ref{alpha}) and $2k-1=\gamma + 1/2~(k=d)$
into (\ref{roots}) yields
\begin{equation}
a_{\pm}=-\frac{1}{2}\left(\gamma + \frac{1}{2}\right) \pm \frac{1}{2}.
\label{3rts}
\end{equation}
The corresponding deformation factor is
\begin{equation}
\chi _{n} = \left(2n + \gamma - \frac{1}{2}\right)
\,\left(2n + \gamma + \frac{3}{2}\right), \label{3def}
\end{equation}
which turns out to be
\begin{equation}
\chi _{n} = E_{n-1}\,E_{n}
\end{equation}
where $E_{n}=E_{n}^{(+)}=E_{n}^{(-)}$.
The structure factor is written as
\begin{equation}
\phi _{n}= n \left(n + \gamma + \frac{1}{2}\right) E_{n}\,E_{n-1}.
\label{3phi}
\end{equation}
Therefore, with $d=(2\gamma +3)/4$, we have
\begin{equation}
\hat{D}_{+}|d, n \rangle =\sqrt{(n+1)(n + \gamma  + 3/2)E_{n}E_{n+1}}\,|d, n+1\rangle
\end{equation}
\begin{equation}
\hat{D}_{-}|d, n \rangle =\sqrt{n(n + \gamma  + 1/2)E_{n}E_{n-1}}\,|d, n-1\rangle
\end{equation}
which are consistent with the SUSY relations (\ref{rai1}) and
(\ref{low1}).

\subsection{Coherent States for the conditionally solvable oscillator}

Utilizing the deformation factor (\ref{3def}) we obtain
\begin{equation}
[\chi _{n}]! = 4^{n}\left(\frac{1}{2}\gamma + \frac{3}{4}\right)_{n}
\,\left(\frac{1}{2}\gamma + \frac{7}{4}\right)_{n} \label{chi3}
\end{equation}
with which we can construct two sets of coherent states as follows.
~\\

\begin{figure}
  \includegraphics[width=500pt]{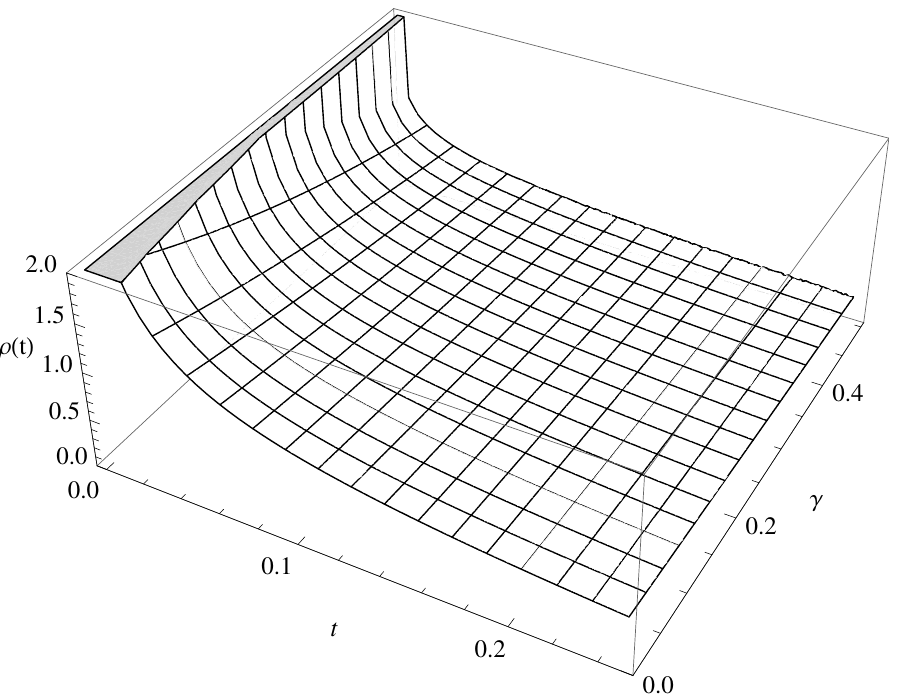}\\
  \caption{The weight function $\rho(t)$ of eq.~(\ref{weight3BG}) for the Barut-Girardello coherent
  states with $t=|\xi|^2$, which is plotted for the allowed range of the characteristic
  parameter $\gamma$ of the conditionally solvable oscillator.}
  \label{Fig1}
\end{figure}

\noindent \underline{\bf Coherent states of the BG-type}: Since the generalized
factorial of the structure factor can be written as
\begin{equation}
[\phi _{n}]!=n!\,(\gamma + 3/2)_{n}\,[\chi _{n}]!, \label{phifac3}
\end{equation}
substitution of (\ref{chi3}) into (\ref{phifac3}) yields
\begin{equation}
[\phi_{n}]! =2^{2n}\,n!\,(\gamma + 3/2)_{n}\,(\gamma /2 + 3/4)_{n}\,
(\gamma /2 + 7/4)_{n}.     \label{phic}
\end{equation}
Inserting (\ref{phic}) into (\ref{BG02}) we have the coherent states for
the cubic algebra of the modified radial oscillator
\begin{equation}
|\xi \rangle = N_{2}^{-1}(|\xi |)\,\sum_{n=0}^{\infty }
\frac{1}{\sqrt{n!\,(\gamma + 3/2)_{n}\, (\gamma /2 +
3/4)_{n}\,(\gamma /2 + 7/4)_{n}}} \left(\frac{\xi}{2} \right)^{n}|\psi _{n}^{(-)}
\rangle   \label{cubBG}
\end{equation}
with the normalization
\begin{equation}
N_{2}^{2}(|\xi |)=\,_{0}F_{3}\,\left(\gamma + 3/2, (2\gamma +3)/4,
(2\gamma + 7)/4; ~|\xi |^{2}/4\right).
\end{equation}
It is apparent that the above coherent states are temporarily stable in
Klauder's sense \cite{JRK} that they evolve with the effective
Hamiltonian,
\begin{equation}
\hat{\cal H}=\frac{1}{2}\omega \hbar\left(\hat{H}_{-} - \gamma -
\frac{3}{2}\right) \label{Heff}
\end{equation}
as
\begin{equation}
e^{-i\hat{\cal H}t/\hbar}|\xi \rangle = |\xi e^{-i\omega t}\rangle .
\end{equation}
The weight function for the resolution of unity (\ref{unity}) is
\begin{eqnarray}
\rho (|\xi |^{2})&=&\left[4\,\Gamma (\gamma + 3/2)\, \Gamma (\gamma /2 +
3/4)\,\Gamma (\gamma /2 + 7/4)\right]^{-1}\, \nonumber \\
&\times & G_{0 \, 4}^{4\, 0}\left(|\xi |^{2}/4
~\left|\,0, \,\gamma + 1/2, \,(2\gamma -1)/4, \,(2\gamma + 3)/4
~\right. \right). \label{weight3BG}
\end{eqnarray}
The coherent states obtained here are basically equivalent to those
proposed by Junker and Roy \cite{JR} if $\varepsilon = -\gamma - 1/2$.
Figure \ref{Fig1} shows
the above weight function for the allowed range of parameter $\gamma$.
~\\

\begin{figure}
  \includegraphics[width=500pt]{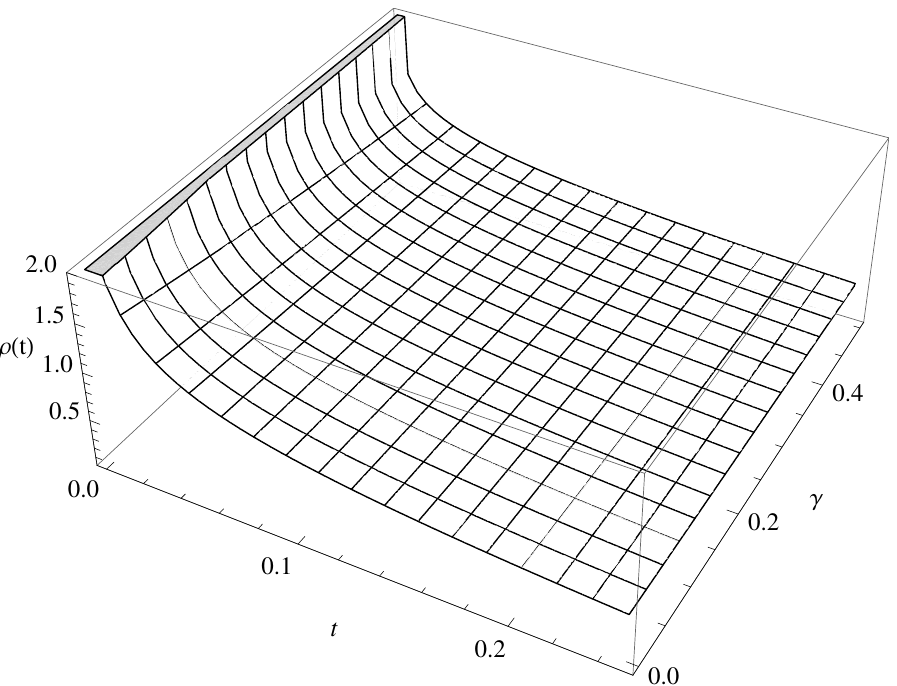}\\
  \caption{The weight function $\rho(t)$ of eq.~(\ref{weight3P}) for the Perelomov type coherent states
  with $t=|\eta|^2$, plotted for the same oscillator.}
  \label{Fig2}
\end{figure}

\noindent \underline{\bf Coherent states of the P-type}: With the same
deformation factor (\ref{chi3}), the coherent states of the P-type for
the cubic case follows from (\ref{CSP2}),
\begin{equation}
|\eta \rangle = N_{2}^{-1}(|\eta |)\,\sum_{n=0}^{\infty
}\left[\frac{(\gamma + 3/2)_{n}}{n!\,(\gamma /2 + 3/4)_{n}(\gamma /2 +
7/4)_{n}} \right]^{1/2}\,\left(\frac{\eta }{2}\right)^{n}|\psi _{n}^{(-)}\rangle,
\label{CSP3}
\end{equation}
with
\begin{equation}
N_{2}^{2}(|\eta |)= \,_{1}F_{2}\left(\gamma + 3/2; \gamma /2 + 3/4,
\gamma /2 + 7/4; |\eta |^{2}/4\right).
\end{equation}
These coherent states are also temporarily stable under the time evolution
with the Hamiltonian $\hat{\cal H}$, that is,
\begin{equation}
e^{-i\hat{\cal H}t/\hbar}|\eta \rangle = |\eta e^{-i\omega t}\rangle .
\end{equation}
The resolution of unity is achieved with the weight function,
\begin{equation}
\rho (|\eta |^{2})=\frac{\Gamma (\gamma + 3/2)}{4
\,\Gamma (\gamma /2 + 3/4)\Gamma (\gamma /2 + 7/4)}\,
G_{1\, 3}^{3\, 0}\left(\frac{|\eta |^{2}}{4}
~\left|\begin{array}{c}\gamma + 1/2 \\ 0, (2\gamma -1)/4, (2\gamma + 3)/4
\end{array}~\right. \right),\label{weight3P}
\end{equation}
which is shown in Figure \ref{Fig2}.
The P-type coherent states are of course different from the BG-type
states.

\section{Concluding remarks}

Extending the deformed algebra $su_{\Phi }(2)$ of Bonatos,
Danskaloyannis and Kolokotronis to $su_{\Phi }(1,1)$ by a simple
analytic continuation, and imposing the polynomial condition on the
structure function, we have proposed a unified way to construct a
discrete set of coherent states of the Barut-Girardello type and of the
Perelomov type for the polynomial $su(1,1)$ algebra.

We have also studied the connection between the cubic algebra $su(1,1)$
and the conditionally solvable oscillator with broken SUSY. We found
that the  eigenstates of the Hamiltonian $\hat{H}_{+}$ in SUSY quantum
mechanics can be identified with a standard basis of the $su(1,1)$
algebra whereas the set of eigenstates of the other partner Hamiltonian
$\hat{H}_{-}$ are identified with a representation space of the cubic
algebra $su_{3}(1,1)$. Then we construct coherent states of the
Barut-Girardello type and of the Perelomov type for the conditionally
solvable system.

The procedure used in the present paper works only for polynomials of
odd degree. In order to accommodate a polynomial algebra of even degree,
such as the quadratic algebra, we have to modify the approach. Although
our consideration is focused on the discrete class, a question remains
open as to whether the same procedure may be extended to a continuous
class in a way similar to that of an earlier work\cite{TI}.

\newpage

\end{document}